# She's Reddit: A source of statistically significant gendered interest information?[1]

Mike Thelwall, Emma Stuart. Statistical Cybermetrics Research Group, University of Wolverhampton, WV1 1LY, UK

Information about gender differences in interests is necessary to disentangle the effects of discrimination and choice when gender inequalities occur, such as in employment. This article assesses gender differences in interests within the popular social news and entertainment site Reddit. A method to detect terms that are statistically significantly used more by males or females in 181 million comments in 100 subreddits shows that gender affects both the selection of subreddits and activities within most of them. The method avoids the hidden gender biases of topic modelling for this task. Although the method reveals statistically significant gender differences in interests for topics that are extensively discussed on Reddit, it cannot give definitive causes, and imitation and sharing within the site mean that additional checking is needed to verify the results. Nevertheless, with care, Reddit can serve as a useful source of insights into gender differences in interests.

## 1 Introduction

Gender inequality is an internationally pervasive problem (UNDP, 2016) with causes that are poorly understood. It is difficult to distinguish between gender differences due to relatively free choice and gender inequalities due to discrimination or social constraints. For example, the substantial gender differences in interest in some toys, books, movies, and TV shows are largely the results of millions of personal decisions that take place within a gendered social context (Simpson, 1996; Cherney & London, 2006; Wühr, Lange & Schwarz, 2017). In the critical area of employment, substantial gender differences in careers are partly due to recruitment decisions and social expectations but are also affected by personal choices (Blau and Kahn, 2000). For example, it has been argued that in the USA women prefer 'people-oriented' jobs, and men prefer 'things-oriented' jobs to some extent (Su, Rounds & Armstrong, 2009) and are differently motivated by the ability of jobs to provide personal or social gains (Diekman, Brown, Johnston, & Clark, 2010). It is difficult to eliminate gender inequality in society if poorly understood gender differences in goals and interests lead to undesired gender imbalances in career choices, social activities and political engagement. Thus, a detailed understanding is needed of gender differences in interests.

Gender differences in interests have been investigated from the perspective of brain functioning (neurology), individual psychology, social psychology, sociology and even through comparisons with non-human animal behaviours (Hines, 2011). None of these perspectives are sufficient on their own, few answers are likely to be definitive and method and data triangulation are essential to give confidence in any hypotheses. One of the problems with understanding gender differences in interests is that there is little public information for even the most obvious cases. For example, systematic data on gender differences in cultural consumption (books, television, films, magazines) is rarely shared with a useful level of detail (Lagaert & Roose, 2015). It is expensive to collect (NEA, 2007) and may be commercially protected. This gap is partly filled by public surveys, such as the National Endowment for the Arts Survey for Public Participation in the Arts, but these necessarily only investigate broad categories and have limited sample sizes. Social web sites give a partial solution to the need for relevant data (Holtz, Kronberger, & Wagner, 2012) by offering public collections of texts discussing many different issues that can be mined for evidence of gender differences in interests or activities (e.g., Walton & Rice, 2013). Few popular sites allow large-scale free data collection, however.

[1] Thelwall, M. & Stuart, E. (in press). She's Reddit: A source of statistically significant gendered interest information? Information Processing & Management.

The social news and entertainment site Reddit has several features that make it a promising source of gender interests for the topics discussed and the demographics that discuss them. In April 2018 it was ranked the 6th most visited website by Alexa.com and 4th in the USA, which accounted for 59% of its visitors[2], with 1.69 billion users accessing the site in March 2018[3]. Reddit allows researchers to mine its data, in contrast to Twitter (limited recent tweets) and Facebook (limited anonymised data from public parts). Topic discussions are also an important part of its appeal, unlike the information dissemination focus of Twitter (Kwak, Lee, Park & Moon, 2010; Chung & Yoon, 2013). These topics may be less constrained than those on friend-based social network sites like Facebook. Reddit user comments can be downloaded through its free API or from a set of public zipfiles. Despite this, there is almost no research into Reddit with a social goal and few studies that investigate gender in the site (see below).

This study investigates Reddit to get evidence of gender differences in topics of interest within this site. It focuses on obtaining statistically significant evidence of gender difference in interests, avoiding prior methods (e.g., topic modeling) that cannot give robust gender difference evidence.

# 2 Background

This review analyses gender differences in conversation topics to illustrate the methods previously used to identify interests and gender differences that are known to exist in some contexts. It also reviews prior research into discussions within bulletin boards and Reddit. The results section contains mini-reviews of gendered interests for the topics found.

## *2.1 Face to face conversation topics*

Informal conversation topics are a potential source of evidence about people's interests. Almost nothing is known about the topics informally spoken about by males and females, however, but there seem to be gender differences in some dimensions, at least for students. Given the wide variety of contexts in which people chat and the likely influence of age, nationality and other demographic information, these results should not be generalised.

An analysis of 261 overheard conversations between University of Michigan students in public spaces in 1990 (Bischoping, 1993) found that females were four times more likely to discuss people of the opposite sex (24% vs. 6%) whereas males were slightly more likely to discuss work and money (43% vs. 38%) or leisure activities (39% vs. 26%). The greater tendency of males to discuss leisure was probably due to more discussions about sports. A small sample investigation of 19 overheard London university cafeteria discussions between students in about 1996 found that females discussed personal relationships more (41% vs. 35%) whereas males spent more time discussing sport/leisure (4.4% vs. 3.7%), work/studies (13% vs. 9%), politics (2.6% vs. 2.2%), and technical/instructional (6.4% vs. 0.8%) topics, whereas there was almost no difference in the amount of discussion about music/culture or personal experiences (Dunbar, Marriott, & Duncan, 1997). In a survey of 515 mainly psychology undergraduates at an American university, females reported discussing people and relationships more often, whereas males discussed movies, music and work relatively more often, although females self-reported discussing most topics more often overall (Sehulster, 2006).

[2] https://www.alexa.com/siteinfo/reddit.com
[3] https://www.statista.com/statistics/443332/reddit-monthly-visitors/

## 2.2 Online conversation topics

The influence of gender may be different for online compared to offline conversation topics because of the nature of the chosen communication medium (e.g., chatroom, social network site), whether the conversation is public or permanent, the temporal dimension (real-time or asynchronous), whether the participants are friends, and the lack of a face-to-face component.

An analysis of a million random Facebook status updates from June 2012 clustered the topics and found that those identifying as female were slightly more likely to mention relationships and personal information, whereas those identifying as male wrote twice as often about sports, politics and religion (Wang, Burke, & Kraut, 2013). A keyword analysis of Facebook messages from 75,000 volunteers found females use terms associated with social processes and home more whereas males were more likely to use terms associated with work and achievements (Schwartz, Eichstaedt, Kern, et al., 2013).

## 2.3 Bulletin boards

Bulletin boards pre-date the web as a mechanism for strangers to exchange ideas about a common topic. Users can select a topic-based forum and post a public message to it, sometimes with multimedia content or links. Bulletin boards often foreground the latest contributions, even if there is a search facility or a list of top posts. Because of their theme-based nature, they have been previously used to investigate public attitudes about individual narrow topics, including from gender perspectives (Gooden & Winefield, 2007; Mo, Malik, & Coulson, 2009). There does not seem to be any systematic research into the gender composition of bulletin board users, perhaps because there are multiple types. They may be obsolete, with users preferring the uniform environment of Reddit.

## 2.4 Reddit

Reddit has offered a public space for discussion since its foundation in 2005. It is like an electronic bulletin board because users can choose a subreddit based on its theme and then read or post comments to discuss the topic. It is unlike friend-based social network sites because friend connections between users are not important. Reddit differs from informal friend-based conversations through being overtly public and not between friends. Nevertheless, these contexts share an informal nature and the ability of the participants to control conversation topics to some extent. Users can comment on posts and both comments and posts can be “upvoted” by members, which makes them more visible, or “downvoted”. Users seem to upvote contributions without reading them, however (Glenski, Pennycuff & Weninger, 2017), so votes are not a reliable measure of content quality.

Subreddits have a defined theme and may self-police through volunteer moderators. New users used to be pre-subscribed to a list of popular subreddits (ExplainLikeImFive, Books, Television, ShowerThoughts, NoSleep, TIFU) but could unsubscribe from them and/or subscribe to others. Presence on this former default list has generated many new subscribers for these communities (Lin, Salehi, Yao, Chen & Bernstein, 2017; see also: Juergens & Stark, 2017). In 2017, Reddit replaced this strategy with a longer list of popular subreddits to help new members find interesting communities (Reddit, 2017). Subreddits can also attract substantial memberships through external publicity (Kiene, Monroy-Hernández, & Hill, 2016).

Prior quantitative Reddit research has exploited the large quantity of public text data to investigate many topics. One study tracked term use in r/depression, r/happy, r/diabetes, and r/ibs, showing that linguistic markers improved over time with participation (Park & Conway, 2017). Several studies have applied topic modelling to identify common topics of discussion in subreddits, including 15,400 r/DarkNetMarkets posts (Porter, 2018) and 1,802 r/STD posts (Nobles, Dreisbach, Keim-Malpass, & Barnes, 2018). Machine learning has also been harnessed to detect text types or relationships within subreddits, such as descriptions of racism (Yang & Counts, 2018), cyberbullying (Rakib & Soon, 2018) and collaboration structures in r/place (Rappaz, Catasta, West, & Aberer, 2018). A few papers have also shown that post popularity in terms of upvotes can be predicted with a reasonable degree of accuracy from user and text properties (Horne, Adali, & Sikdar, 2017; Lim, Carman, & Wong, 2017).

Content or thematic analysis are suitable for analysing Reddit posts. For example, an assessment of comments posted to four question-based subreddits in 2015/2016 found that explanations with disagreements accounted for about half of the r/ask_Politics comments, whereas neutral explanations accounted for about half of the comments in the remaining subreddits (askAcademia, askScience, askHistorians) (Kumar, Gruzd, Haythornthwaite, Gilbert, Esteve del Valle & Paulin, 2018). Less common activities included socialising, information seeking and providing resources. Pro-eating disorder communities on Reddit have also been investigated, showing that some participants supported problematic behaviours (Sowles et al., 2018). An analysis of the r/Gout subreddit, found most questions to be about diagnosis (Derksen, Serlachius, Petrie, & Dalbeth, 2017).

There are many small-scale qualitative analyses of subreddits. These include r/ExplainLikeImFive (Pflugfelder, 2017) and (partly quantitative) r/RoastMe (Kasunic & Kaufman, 2018). Not all analyses of Reddit have operated at the level of subreddits. One study of vaping identified relevant discussion threads from the whole site through a set of curated queries (Sharma, Wigginton, Meurk, Ford, & Gartner, 2016). Another identified 38 small dermatology-related subreddits to analyse using a set of keyword queries (Buntinx-Krieg, Caravaglio, Domozych, & Dellavalle, 2017). Finally, some researchers have recruited participants for questionnaires or interview studies unrelated to the site from subreddits (Jamnik & Lane, 2017).

## *2.5 Gender on Reddit*

There is little prior research on Reddit that explores the content posted or gender. In August 2018 the Scopus search TITLE-ABS-KEY((male OR female OR gender) AND (reddit OR redditor OR subreddit)) returned 27 matches. None of these used a quantitative approach to investigate gender differences in interests. For example, one quantitative study tracked gendered terms applied to US presidential candidates over time (Hale & Grabe, 2018) another assessed whether user gender associated with self-declared weight loss in r/LoseIt (Pappa, Cunha, Bicalho, Ribeiro, Silva, Meira & Beleigoli, 2017) and a third focused on detecting user genders (Fabian, Baumann, & Keil, 2015). Other studies used qualitative methods and did not focus on binary gender differences (Brennan, Swartout, Cook, & Parrott, 2018; Chang-Kredl & Colannino, 2017; Darwin, 2017; Farber, 2017; Marwick, 2017; Massanari, 2017; Smith, 2015).

The most relevant quantitative study compared the topics discussed in the r/Parenting, r/Mommit, and r/Daddit subreddits between 2008 and 2016 (Ammari, Schoenebeck, & Romero, 2018). It applied topic modelling to the subreddits collectively and separately and assessed the extent to which each of the topics occurred in the male, female and neutral subreddits. Gender was not inferred for individual users. The study found topics that were more discussed in the male parenting board (e.g., Halloween Costumes) or the female parenting board (e.g., Child Weight Gain). It relies upon gendered subreddits, limiting its generalisability. It also did not filter for multiple posts from the same redditor, and in some cases (e.g., Discipline, Teen/Sex Talk) a topic was discussed much more in the neutral board than for either gender, which seems counter-intuitive.

One other study has investigated gender on Reddit. A survey of the atheism subreddit in 2011 through posts soliciting age and gender information obtained responses from 734 users. Males dominated this subreddit (64% male, 17% female, 19% no answer), as did younger people (only 8% older than 35). Females wrote longer posts and older people's posts tended to be upvoted more (Finlay, 2014). The lack of site-wide Reddit analyses is unfortunate given the potential of the site to reveal gendered interests in topics.

## 2.6 Quantitative methods to detect gender differences in interests

Many different methods have been used to detect gender differences in social media. A logical approach is topic modelling (related: Van Oerle, Mahr, & Lievens, 2016) to detect topics within a collection of posts and then to assess the extent to which male and female posts fall within these topics. The results are biased by known stylistic differences between genders that might affect both the topics generated and the extent to which each gender associates with a topic. For example, if OMG is part of a topic then this will tend to associate the topic with females, irrespective of the other terms used. Topic modelling can also be used to cluster users rather than their posts (Qiu & Shen, 2017).

An investigation of international and gender differences in how mental illness was self-disclosed on Twitter analysed Tweets with a relevant search phrase in March 2015, inferring genders from usernames, and gathering the most recent 3200 tweets for each user, discarding any with less than 75% English tweets (De Choudhury, Sharma, Logar, Eekhout, & Nielsen, 2017). Machine learning identified Tweeters with genuine mental health disclosures, harnessing posts from the r/depression, r/mentalhealth, and r/SuicideWatch subreddits to help. Topic modelling detected the main topics of the Twitter users and linguistic markers were extracted from their tweets. Gender differences in these were detected by averaging male and female user contributions. For example, males were less likely to discuss the need for social support. Topic modelling was also used for contributions to Tianmijiayuan, an online discussion forum for diabetes, and gender differences found in the number of posts assigned to each topic (Liu, Sun, & Li, 2018). It has also been applied to Sina Weibo posts from Chinese tourists in Japan (Xia, Song, Huang, Miyazawa, Fan, Jiang, & Shibasaki, 2017).

Some analyses have used pre-defined lexicons and counted matching texts for one or more genders. One study detected gender-fluid individuals from their tweets and identified potential mental health issues with a lexicon matching approach (Zhao, Guo, He, Huo, Wu, Yang, & Bian, 2018). Another matched lexicons of terms in a curated collection of tweets about gender-based violence to assess gender differences for a set of topics, including humour and soccer (Purohit, Banerjee, Hampton, Shalin, Bhandutia, & Sheth, 2016).

A different approach was used in an analysis of gender differences in climate communication on Twitter. Gender differences were found in the proportions of males and females mentioning any username or hashtag and then those with substantial gender differences were manually classified for user stance or hashtag topic (Holmberg & Hellsten, 2015). This has similarities to the current paper but does not safeguard familywise error rates in multiple statistical tests.

A hybrid method to detect gendered interests on Twitter associated users with Wikipedia pages when the names matched (e.g., @BritneySpears and "Britney Spears") and then identified gendered interests by assessing the gender proportions of followers of the matched Twitter users (Faralli, Stilo, & Velardi, 2015). This identifies gendered levels of interest in popular individuals rather than topics, however.

Content analysis has also been used to detect gender differences in political topics of interest on Twitter for US House representatives, showing that female issues are not the only topic favoured by women (Evans, 2016). This strategy is only practical for small samples of users, however.

# 3 Research questions

This paper assesses whether a quantitative analysis of Reddit can give statistical evidence of gender differences in interests in terms of (a) participation rates and (b) contributions. The first research question does not assess statistical significance because participation rates are affected by gender differences in the uptake of Reddit and the existence of competing subreddits with similar themes.

1. Does the proportion of female subreddit participants vary substantially by topic?
2. Are there statistically significant gender differences in contributions to individual subreddits?
3. Can the answers to the above questions illuminate underlying gender differences in interests?

# 4 Methods

The research design was to download all comments from Reddit, identify the gender of the commenters through their names, select a comment from each commenter at random for each subreddit-month, and then, for the top 100 subreddits, identify the proportion of female commenters (RQ1) and use a word frequency analysis to identify terms that associate with males and females (RQ2). The final stage was to critically analyse the results from the perspective of underlying gender interests (RQ3). The steps are explained in detail and justified below.

## *4.1 Data*

Comments posted to Reddit were downloaded in zipfiles collected by Jason M. Baumgartner and posted to https://files.pushshift.io/reddit/comments/. This is a large but not comprehensive collection of Reddit posts, with some temporal gaps (Gaffney & Matias, 2018). The files were downloaded in March 2018, covering the period December 2005 to February 2018. This produced 3,683,577,011 non-empty, non-deleted comments that did not consist of an URL alone. These are comments made in response to prior posts to a subreddit. The original posts might be questions or URLs, although longer posts (e.g., short stories) were allowed in some subreddits. The number of comments per month was low from 2005 until 2010 (e.g., 2,843,316 in January 2010) and then grew approximately linearly until February 2018 (80,397,389 comments).

Reddit members do not register a gender when joining but it is sometimes possible to guess a male or female gender from their usernames. Usernames were matched against a list of 4772 first names from the most popular 10,000 that are at least 90% male or female from the United States 1990 census[4]. Users with non-matching names were left as ungendered. First names were extracted by splitting usernames using camel case (MikeThelwall -> Mike) or at the first number (Mike33 -> Mike). Although gender can be inferred from users' posts (Fabian, Baumann, & Keil, 2015), this would skew the text analysis results, interfering with the research questions. The first name gender detection method makes mistakes because of people with unusual gender first names, minority cultures with differing gender associations for the same name (e.g., Italian) and short form names that swap gender (e.g., Ali for Alison).

[4] https://www.census.gov/topics/population/genealogy/data/1990_census/1990_census_namefiles.html

It may also be more accurate for one gender, biasing the results. Based on an analysis of YouTube, which has a similar demographic, the results probably over-represent the presence of males by up to 3.5% (doi:0.6084/m9.figshare.5688622). The errors are likely to cancel out and not affect RQ1. They will weaken the statistical power for the analyses in RQ2 because some comments will be associated with the wrong gender. Assuming the gender swapping is not extensive or systematic, this will not invalidate the RQ2 statistical results.

After discarding comments from users for which genders cannot be inferred, there were 180,545,882 non-duplicate comments from (inferred) males or females, which is 4.9% of the set of non-deleted comments. This low percentage is due to usernames being creative aliases that typically do not reflect the offline names of the individuals. To avoid the results being dominated by individual prolific posters, only one comment was allowed per user, per subreddit, per month (or one per user, per month for the word frequency analyses of all Reddit: see below). For users posting more than once in a month, a post was selected using a random number generator. Allowing one post per month is a compromise because allowing only one post per user for the full dataset would have given too much weight to casual users that posted once in comparison to the dedicated users that are presumably the heart of Reddit. Excluding ungendered comments, there were 152,532 subreddits with at least one comment from an author classified as male or female. Counting a maximum of one comment per gendered user per month, there were 36,068,350 comments altogether, 24.2% of which were from authors with female names.

After removing non-gendered and duplicate comments, the 100 subreddits with the most comments were selected for analysis. These 100 subreddits accounted for half (15,361,819; 42.6%) of all comments, with a quarter (3,532,206; 23.0%) from authors classified as female. Alternative approaches to detect topics of interest include clustering algorithms and topic modelling but all text-based approaches would generate topics contaminated by gender differences in writing styles (Biber & Conrad, 2001; Iqbal, Ashraf, & Nawab, 2015). Using the pre-defined and user-selected subreddits avoids this problem, is fully transparent and allows an emphasis on the user's decision about where in Reddit to contribute. Using the subreddit as a unit of analysis also avoids problems in RQ2 caused by differing post lengths and cultures between subreddits.

## 4.2 Analyses

The subreddits were examined individually and grouped subjectively into major themes to identify broad topic trends to help address RQ1 (although they are not strictly necessary). Commenter gender was not used to decide these themes. In some cases, subreddit topics were obvious from their titles (e.g., r/battlefield3 for the computer game Battlefield 3) and in others (e.g., r/IAmA, r/TwoXChromosomes), subreddit self-descriptions explained the permitted contribution types or moderation policy. Comments were read to detect the dominant topics of general subreddits to check that the ostensible purpose did not mask different practices. The major themes detected by subjectively clustering the subreddit topics nevertheless represent one of many potentially meaningful groupings. Whilst a clustering algorithm or topic modelling could have been used, this could interfere with the term-based analysis of RQ2 since gender influences writing style (Biber & Conrad, 2001) and participants often used strongly gendered terms unrelated to the topic in their comments, such as *husband, boyfriend, girlfriend* and *wife*. Any algorithm would also cluster by topic and/or style whereas it is more useful to cluster by purpose, since this seems to be more fundamental to the site. The final clusters, for example, grouped very different format subreddits together because of a similar purpose (e.g., r/AdviceAnimals pictures and r/NotTheOnion for political humour). A hierarchical cluster analysis (using TFIDF similarity between the top 100 subreddits) was also constructed to investigate if automatic clustering could generate meaningful clusters but the results were not helpful. For example, r/BlackPeopleTwitter (comedy) was clustered with r/Music and r/HipHopHeads, whereas r/Jokes and r/Funny were not clustered together.

It would be possible to group together similar subreddits for a more powerful RQ2 statistical analysis. For example, r/Gaming and r/Games employ similar terms (TF-IDF cosine similarity 0.748). This was not done in the current paper to give a clear focus on subreddits. Even ostensibly similar subreddits could have different term usage (e.g., r/WTF and r/TIFU have a cosine similarity of only 0.168) and different subreddits can be textually similar (e.g., the book series: r/ASOIAF and TV series based on it r/GameOfThrones have a TF-IDF cosine similarity of 0.891) presumably mainly due to mentions of made-up characters and place names in both.

For RQ2, a standard approach to detect gender differences in sets of texts is to apply topic modelling to each subreddit and identify gendered levels of contribution to the topics discovered (De Choudhury, Sharma, Logar, Eekhout, & Nielsen, 2017; Liu, Sun, & Li, 2018; Xia, Song, Huang, Miyazawa, Fan, Jiang, & Shibasaki, 2017). As mentioned above and discussed in more detail below, this is undesirable because gender differences in linguistic styles (Biber & Conrad, 2001) contaminate the results. The most commonly used Latent Dirichlet Allocation (LDA) algorithm exploits a matrix factorisation approach to build topics from words that frequently co-occur (O'Callaghan, Greene, Carthy, & Cunningham, 2015). Since females are more likely to use explicitly positive language (Thelwall, Wilkinson, & Uppal, 2010), any topic that tends to be discussed positively would have a female bias in its contributions. A positive topic with no underlying gender differences would thus appear to be female-gendered because positive words would contribute to the topic and females author more positive words. Any positive topic extracted by topic modelling is therefore inherently gendered. Other topics can also be accidentally gendered due to natural gender differences in term uses. For example, females are more likely to use terms like boyfriend and husband whereas males are more likely to use girlfriend and wife. If there are topics in which only one gender occasionally uses these terms then this will give a gender association to that topic from the other people using the same words (e.g., wife/husband) in other topics.

Instead, within each subreddit words were identified that were used relatively often by one gender compared to the other. For this, each word occurring in each subreddit (n=2,416,953 for the whole project) was checked (using the word association option in the software Mozdeh: Thelwall, 2009) to see whether it occurred disproportionately often for one gender, using a 2x2 chi-squared association test at the 0.05 level. This tests for an underlying tendency for one gender to use the term in the subreddit more often than the other. Since there was a separate test for each term and therefore millions of tests for each subreddit, many false positive results would be produced. The Benjamini-Hochberg (Benjamini & Hochberg, 1995) procedure was used to control the false discovery rate for these tests at the 5% level, although this is a heuristic because the word frequency data is not independent. It is not independent because users are allowed multiple posts in different months, users may quote or imitate each other, and multiple replies to the same original post are more likely to use similar terms. Nevertheless, this procedure gives a list of terms that may be indicative of an aspect of the theme of the subreddit. To increase statistical power, words that were too rare to have any chance of generating a positive result (i.e., if the word was only used by the minority gender then a chi-squared significance test would still accept the null hypothesis at the uncorrected 5% level) were discarded before the Benjamini-Hochberg calculations.

To illustrate the method, in the Gaming subreddit, 0.55% of female-authored comments contained the term *girl*, in comparison to 0.33% of male-authored comments in the same subreddit. The chi-squared value for this, 111.7, falls within the set judged to be statistically significant by the Benjamini-Hochberg procedure, so the hypothesis that there is a gender difference in the use of the term *girl* for comments in the Gaming subreddit was retained.

For each subreddit, the list of terms judged statistically significant was investigated qualitatively to check whether any were topic-based rather than stylistic. Gendered terms were characterised as stylistic if they were generic (e.g., *and*), expressed sentiment (expressing explicit sentiment online is a female-associated trait: Thelwall, Wilkinson, & Uppal, 2010) or discussed personal relationships as context for a comment. Two of the most gendered terms, *husband* and *boyfriend*, fell into the latter category. Male-associated swearing was characterised as stylistic (Thelwall, 2008), as were the terms *wife* and *girlfriend*. A full list of the significant terms for each subreddit is available online at doi:10.6084/m9.figshare.6470930. The purpose of this exercise was to be able to draw stronger conclusions when gendered topic differences are present. The exact number of topic-based terms in each subreddit is not relevant so just their presence is reported for each subreddit.

A critical analysis of the RQ1 and RQ2 results from the perspective of wider gender issues was used to address RQ3 and is reported in the discussion section for convenience.

# 5 Results

The subreddits grouped into (overlapping) themes are discussed separately below in decreasing order of popularity for the most popular subreddit in each one. The proportion of female commenters is reported in a table (RQ1) together with the number of terms for which a gender difference was found (RQ2). Each subsection also contains a brief description of the theme and gender differences between the subreddits within the theme (RQ1). Some include a brief description of the topic-based terms found for one of the subreddits (RQ2).

## 5.1 General interest

Many subreddits have an information format (e.g., images, questions) rather than a topic. Asking and answering questions is a common human activity and there are popular websites for this, such as answers.yahoo.com. General information subreddits may function as a form of light entertainment for browsing interesting posts. Users can be more active by writing comments or posts or by rating others' contributions. Since posts and comments can be upvoted, participation can be a game-like challenge to create popular content.

*Gendered participation rates*: Since the subreddit topics in this category are general, gendered participation rates reveal nothing about topics, but suggest that engaging with light entertainment in the Reddit format is a male-associated trait.

*Gendered participation differences*: All subreddits except r/InterestingAsFuck reveal gendered topic interests. The generic question-based r/AskReddit is the most popular subreddit. Female comments are more likely to be about relationships (e.g., topic terms included *mom, family, baby*) whereas male comments and questions are more likely to be about games and sports (topic terms included *game, team, player, win, league, play*).

**Table 1**. General subreddits in the top 100, the number of gendered comments (max. one per user per month) and the proportion of gendered comments that are female-authored. Subreddits are ordered by increasing female share. Descriptions are partly extracted from subreddit home pages.

| Rank | Subreddit | Comm. | Fem. | Description | Gendered terms* |
|---|---|---|---|---|---|
| 5 | Videos | 528169 | 19.7% | Video content of all kinds | T,S: 57 cat |
| 7 | TodayILearned | 485965 | 21.5% | Facts that you just found out | T,S: 116 parent |
| 12 | Gifs | 316892 | 21.7% | Interesting or funny gifs | T,S: 35 kitty |
| 64 | InterestingAsFuck | 64683 | 21.8% | Anything interesting | S: 1 |
| 3 | Pics | 851012 | 23.8% | Share photographs and pictures | T,S: 244 mom |
| 19 | MildlyInteresting | 192183 | 24.2% | Mildly interesting photographs | T,S: 19 cat |
| 1 | AskReddit | 2030184 | 27.2% | Questions | T,S: 1599 parent |

*T=Statistically significant evidence of topic differences between genders; S=Statistically significant evidence of style differences between genders. An example gendered topic term is given, when available.

## 5.2 Humour

Fourteen subreddits deal with humour. Online humour has a long tradition, including textual jokes (Shifman, Levy, & Thelwall, 2014), funny pictures (Shifman, 2014), animated pictures (gifs) (Bakhshi, Shamma, Kennedy, Song, de Juan & Kaye, 2016) and videos (Shifman, 2012).

*Gendered participation rates*: Humour-based subreddits are all strongly male oriented, although there are differences in the size of the female minority. Tasteless humour is particularly male-oriented, in contrast to r/Facepalm (things that are stupid). Taken together, the results suggest that browsing Reddit for amusing content is a strongly male-oriented light entertainment activity.

*Gendered participation differences*: All except three subreddits reveal gendered topic interests, with some narrower topics within others. It is surprising that no gender differences were found for (textual) r/Jokes but many were found for r/Funny. This is partly because r/Funny has more comments, making the statistics more powerful. An example of a female-oriented topic term is *pregnant* for r/Facepalm – mainly related to situations where a woman has done something inappropriate for pregnancy. Presumably women can relate to such situations better than men and their potential for humour partly stems from fear of making a mistake when adjusting to a significant life event.

Table 2. Humour themed subreddits.

| Rank | Subreddit | Comm. | Fem. | Description | Gendered terms* |
|---|---|---|---|---|---|
| 93 | ImGoingToHellForThis | 44881 | 19.2% | Tasteless dark, offensive, & twisted humour | 0 |
| 97 | me_irl | 42534 | 19.8% | Selfies of the soul | 0 |
| 80 | reactiongifs | 50817 | 20.4% | Physical or emotional response in an animated gif | T,S: 6 mom |
| 39 | Jokes | 85785 | 21.4% | Text format jokes | 0 |
| 59 | BlackPeopleTwitter | 67491 | 21.8% | Screenshots of Black people being hilarious and insightful on social media | T,S: 5 submitting |
| 37 | NotTheOnion | 87128 | 23.0% | Ridiculous true stories | T,S: 8 mom |
| 2 | Funny | 879047 | 23.5% | Attempt humour | T,S: 164 hair |
| 8 | AdviceAnimals | 403858 | 24.0% | Two line joke over an image of a character, generally an animal | T,S: 286 married |
| 18 | Showerthoughts | 192838 | 24.0% | The amusing/interesting within the mundane | T,S: 25 sister |
| 6 | WTF | 528169 | 24.4% | Things that make you say WTF | T,S: 181 vet |
| 60 | CringePics | 67437 | 25.7% | Embarrassing interactions; often message screenshots | T,S: 6 friend |
| 49 | fffffffuuuuuuuuuuuu | 75177 | 26.3% | A subreddit for rage comics | T,S: 10 girl |
| 33 | TIFU | 93158 | 26.6% | Today I Fucked Up | T,S: 50 tampon |
| 96 | Facepalm | 44051 | 26.8% | Posts should cause viewers to facepalm | T,S: 9 pregnant |

*T=Statistically significant evidence of topic differences between genders; S=Statistically significant evidence of style differences between genders. An example gendered topic term is given, when available.

## *5.3 Gaming*

Playing computer games is a male associated activity, although this may be mainly due to a greater male preference for shooting and (mainly fighting-based) role-playing games (Rehbein, Staudt, Hanslmaier & Kliem, 2016). Greater male enjoyment of violent games may be due to a combination of ethical and preference differences (Hartmann, Möller, & Krause, 2015). Males tend to enjoy all key components of the computer gaming experience slightly more than do females (Scharkow, Festl, Vogelgesang & Quandt, 2015). Female participation within gaming has also been hindered by Gamergate, an online campaign in 2014, which culminated in many females involved within gaming leaving the industry due to harassment from 'young male gamers' (Hathaway, 2014).

*Gendered participation rates*: All game-based subreddits are heavily male dominated, confirming that computer game playing and discussing computer game playing online are male-associated. Differences between games suggest that there are gender variations in the uptake of some ostensibly similar games. In support of this, a much less male-oriented game subreddit than those in Table 3 is r/FFXIV (rank 171, 38.8% female), for the massively multiplayer online role-playing game Final Fantasy XIV, released in 2010. One commentator has claimed that the game has a relatively high ratio of females for the game type but is still male dominated (3 females for every 7 males)[5]. Similar results were obtained from an informal poll[6]. Final Fantasy XIV has a more sophisticated take on gender than normal for the genre, whilst not avoiding sexual content[7]. For example, although it follows genre by having skimpy outfits for females, it breaks genre with skimpy outfits for males[8]. Other similar games, such as World of Warcraft, have social environments that may be uncomfortable for many females (Nardi, 2010).

*Gendered participation differences*: Game subreddit discussions reveal few insights into participants' gendered interests. The main exception, r/Gaming, revealed greater female interest in the Sim game series and gender issues within gaming (topic terms: girl, woman, women).

Table 3. Computer game subreddits.

| Rank | Subreddit | Comm. | Fem. | Description | Gendered terms* |
|---|---|---|---|---|---|
| 61 | XboxOne | 66554 | 15.1% | Related to Xbox One | S: 2 |
| 65 | PS4 | 64401 | 15.5% | Related to PlayStation 4 | 0 |
| 98 | Fallout | 41956 | 17.0% | Single player action role-playing video games | S: 1 |
| 32 | Games | 94013 | 17.6% | Computer games or the game industry | T: 1 problematic |
| 56 | Minecraft | 72260 | 18.0% | Sandbox video game | S: 3 |
| 4 | Gaming | 560440 | 18.2% | Gaming-related | T,S: 18 girl |
| 54 | DestinyTheGame | 72654 | 18.4% | Online multiplayer first-person shooter video game | S: 2 |
| 95 | MagicTCG | 44405 | 19.3% | Single/multiplayer video game based on the popular collectible card game Magic: The Gathering | T,S: 2 assistant |
| 94 | Skyrim | 44553 | 19.5% | Open world single player action role-playing video game | S: 1 |
| 41 | GlobalOffensive | 85221 | 20.3% | Multiplayer first-person shooter video game | 0 |
| 62 | Hearthstone | 66389 | 21.0% | Online collectible card 2-player battle fantasy video game | 1 |
| 40 | Overwatch | 85299 | 23.4% | Team-based multiplayer first-person shooter video game | T: 2 mercy |
| 68 | WoW | 57617 | 23.8% | Massively multiplayer online role-playing game (World of Warcraft) | S: 2 |
| 16 | LeagueOfLegends | 237864 | 24.4% | Multiplayer online battle arena video game | T,S: 4 female |
| 44 | DotA2 | 82440 | 24.8% | Multiplayer online battle arena video game | 0 |

*T=Statistically significant evidence of topic differences between genders; S=Statistically significant evidence of style differences between genders. An example gendered topic term is given, when available.

[5] https://www.reddit.com/r/ffxiv/comments/6orjni/ff14_player_bases_gender_ratio_according_to_yoship/
[6] https://www.reddit.com/r/ffxiv/comments/4kf1ps/gender_and_roll_poll_results/
[7] https://www.reddit.com/r/ffxiv/comments/7ut01f/ffxivs_often_overt_sexual_themes_are/
[8] https://www.reddit.com/r/ffxiv/comments/888gee/someone_explain_the_reason_for_the_obvious/

## 5.4 News/Politics/Religion

Interest and engagement in politics is a male associated trait (Verba, Burns, & Schlozman, 1997), including within Facebook (Brandtzaeg, 2017). Atheism is political in the sense that it is arguing against organised religion, and is more common amongst males (Brewster, 2013). Males are also more interested in news, especially in countries where females are less economically active (Benesch, 2012).

*Gendered participation rates*: The news and politics subreddits have similar gender compositions as each other and to the Reddit average, reflecting both the male domination of Reddit and greater male interest in these topics.

*Gendered participation differences*: Four of the subreddits reveal gendered topic differences. For r/Atheism, female-associated terms indicated relationships (topic terms: daughter, parents, family, mom, friend, relationship, raised), and females (her, she, women). For r/News, female-associated topics included health (health, nurse, doctor, medical), rape, and teachers.

Table 4. News, politics or religion subreddits.

| Rank | Subreddit | **Comm.** | **Fem.** | Description | **Gendered terms*** |
|---|---|---|---|---|---|
| 13 | Politics | 316089 | 20.5% | U.S. Politics | T,S: 18 birth |
| 48 | The_Donald | 76811 | 20.7% | For serious supporters of President Trump | S: 2 |
| 90 | Conspiracy | 45953 | 20.7% | Conspiracy theories | 0 |
| 9 | WorldNews | 374686 | 20.7% | Major news from around the world except US | T,S: 41 victim |
| 83 | Canada | 48836 | 21.7% | Canadian news | S: 1 |
| 23 | Atheism | 157483 | 22.1% | Atheism, agnosticism and secular living | T,S: 33 derailing |
| 15 | News | 285467 | 22.6% | News not politics | T,S: 79 detox |

*T=Statistically significant evidence of topic differences between genders; S=Statistically significant evidence of style differences between genders. An example gendered topic term is given, when available.

## 5.5 Education and science

There is approximate gender parity at all levels of education in the USA (e.g., https://www.dol.gov/wb/stats/NEWSTATS/latest/demographics.htm), although there are large gender differences in academic subject preferences. Females are less likely to study degrees in engineering, mathematics, computer science, and physics (Hyde, 2014) in comparison to health care, elementary education and domestic sphere subjects (Tellhed, Bäckström, & Björklund, 2017).

*Gendered participation rates*: Scientific and educational subreddits have similar gender compositions, suggesting a greater male interest in these topics despite gender parity in education. Space is a traditionally male topic, as is data from the perspective of computing, maths and science, but the male domination of the other areas is surprising. It is possible that there are other forums online that are more natural for these, so participants could be people that have joined Reddit for other topics.

*Gendered participation differences*: Four out of six subreddits reveal gendered topic differences. For r/Science and r/ExplainLikeImFive, there is a female interest in health issues, for example, and for r/IAmA there is a female interest in books and school. Note that r/IAmA could have been listed as a general interest subreddit but its most popular comments seemed to have educational rather than curiosity function, informing about aspects of jobs or careers.

Table 5 Educational or scientific subreddits.

| Rank | Subreddit | Comm. | Fem. | Description | Gendered terms* |
|---|---|---|---|---|---|
| 81 | Space | 50005 | 18.7% | Astrophysics, Cosmology, Space Exploration, Planetary Science, Astrobiology | S: 1 |
| 73 | Futurology | 55810 | 20.0% | Evidence-based speculation about humanity, technology, and civilization | 0 |
| 85 | DataIsBeautiful | 48008 | 21.0% | Graphs, charts, maps, etc. | T,S: 3 women |
| 29 | Science | 111126 | 21.7% | Links to published peer-reviewed research or media summaries | T,S: 39 doctor |
| 25 | ExplainLikeImFive | 137395 | 23.1% | Academic questions answered in simple language | T,S: 41 pain |
| 10 | IAmA | 352255 | 24.6% | Ask me anything | T,S: 321 book |

*T=Statistically significant evidence of topic differences between genders; S=Statistically significant evidence of style differences between genders. An example gendered topic term is given, when available.

## 5.6 Mass entertainment: Music/TV/movies/books

Popular music is somewhat male-oriented, at least for public performance, with male artists outnumbering female artists in the USA by 3.5 to 1 (Smith, Choueiti, & Pieper, 2018). In contrast, listening to music seems to be an almost universal human pastime. Nevertheless, youth music preferences and subcultures evolve dramatically over time and encompass substantial gender differences (Christenson & Peterson, 1988). Whilst the existence of music subcultures presupposes a degree of communication about music, no previous study seems to have compared the extent to which males and females discuss it. Males in the USA watch more TV[9] but TV does not seem to have been investigated as a topic of conversation. Presumably, it would be difficult to separate TV watching from discussions of the topics watched in some cases, such as news and sport, but not in others, such as drama. This would cause a natural bias in any study of gender differences in conversations. In terms of movie and TV genres, one commercial source suggests that in the USA, females are more likely to like romances, dramas, kids & family, whereas males are more likely to enjoy science fiction and action[10].

In the USA and internationally, girls read better than boys (Loveless, 2015), perhaps because they are encouraged to read more for pleasure (Clark, Osborne & Akerman, 2008). In the USA, females are twice as likely to read fiction, partly due to childhood encouragement (Tepper, 2000). In 2002 in the USA, 55% of adult females read fiction compared to only 38% of adult males (Bradshaw & Nichols, 2004). In the USA, males may write more published fiction than females. For HathiTrust books (from research libraries), the proportion of female authors of English language fiction increased from 25% in 1970 to 40% in 2009 (Underwood, Bamman, & Lee, 2018). These libraries seem likely to hold a low proportion of genre fiction, such as category romances and young adult, so the overall proportion of female authors is likely to be higher.

[9] https://techcrunch.com/2012/10/05/nielsen-gaming-tv-console/
[10] https://www.statista.com/statistics/254115/favorite-movie-genres-in-the-us/

*Gendered participation rates*: The mass entertainment subreddits have differing gender compositions but all are at least two thirds male. Hiphop is a male dominated music culture and Star Wars has a high proportion of male fans (Collins, Hand, & Linnell, 2008), presumably for its science fiction, space and fighting themes. The other subreddit topics seem to be more gender neutral, suggesting that the gender imbalance is inherited from the male dominance of Reddit. Some popular subreddits outside the top 100 in this category have mostly commenters identified as female, including r/RuPaulsDragRace (rank 269, 51.7% female) and r/GilmoreGirls (rank 2781, 76.7% female), but these are rare exceptions.

*Gendered participation differences*: Most of the subreddits reveal a few gendered topic differences. For r/Books, female associated terms include genres (romance), authors ([Jane] Austen), and books ([The] handmaid's [tale]) as well as associated activities (library, re-read).

Table 6. Mass market music, television, movie and books subreddits.

| **Rank** | **Subreddit** | **Comm.** | **Fem.** | **Description** | **Gendered terms*** |
|---|---|---|---|---|---|
| 52 | HiphopHeads | 72922 | 14.8% | Hip hop mixtapes, videos, news | 0 |
| 67 | StarWars | 60307 | 17.8% | Anything Star Wars | S: 1 |
| 11 | Movies | 342581 | 18.9% | News and discussion of major motion pictures | T,S: 27 princess |
| 28 | Television | 119296 | 20.8% | TV shows | T,S: 8 Buffy |
| 17 | Music | 201001 | 21.2% | Artist, title and genre | T,S: 30 Brendon |
| 88 | ASOIAF | 46190 | 22.8% | A song of Ice and Fire fantasy book series | T,S: 6 Sansa |
| 57 | GameOfThrones | 69558 | 22.9% | About the TV series | T,S: 8 hair |
| 43 | Pokemon | 82740 | 23.8% | Pokémon TV shows, video games, toys, trading cards | T,S: 4 artist |
| 71 | Anime | 57100 | 28.1% | Anime TV series, comics and fan art | S; 1 |
| 34 | Books | 92631 | 30.1% | In-depth discussion about books, authors, genres or publishing | T,S: 58 McCafferty |

*T=Statistically significant evidence of topic differences between genders; S=Statistically significant evidence of style differences between genders. An example gendered topic term is given, when available.

## 5.7 Technology/computers

All the technology and computing subreddits are even more male dominated than Reddit overall, reflecting a strong male interest in these topics (Su, Rounds, & Armstrong, 2009). The results give few insights into gender differences within each subreddit.

Table 7. Technology or computing subreddits.

| Rank | Subreddit | Comm. | Fem. | Description | Gendered terms* |
|---|---|---|---|---|---|
| 89 | Programming | 45961 | 16.0% | Related to computer code | 0 |
| 47 | Android | 76876 | 16.0% | About the mobile operating system developed by Google | S: 1 |
| 82 | Apple | 49171 | 16.0% | Apple news, rumours, and discussions | S: 1 |
| 22 | PCmasterrace | 175957 | 16.6% | Community of PC enthusiasts | T,S: 2 terramuncher |
| 63 | BuildaPC | 66332 | 17.0% | Help building a computer | 0 |
| 21 | Technology | 176084 | 18.2% | Latest technology developments, happenings and curiosities | T,S: 3 poker |

*T=Statistically significant evidence of topic differences between genders; S=Statistically significant evidence of style differences between genders. An example gendered topic term is given, when available.

## 5.8 Sport

Males have a deeper interest in competitive sport than females (Gantz & Wenner, 1991) and are more likely to talk about it (Dietz-Uhler, Harrick, End & Jacquemotte, 2000; James & Ridinger, 2002), to read about sports (Hughes-Hassell & Rodge, 2007) and to follow sports news (Taylor, Funk, & Craighill, 2006). Males are far more likely to play fantasy sports than females (9% vs. 1% in: Taylor, Funk, & Craighill, 2006). Gyms seem to have a male culture (Pridgeon & Grogan, 2012), and women may employ other methods to exercise, such as aerobics (e.g., 51% more females in Australia participated in aerobics/fitness activities in 2002: ABS, 2003) and perhaps home fitness videos.

*Gendered participation rates*: All the sports subreddits are strongly male dominated, reflecting male interest in sports. The least male dominated, fitness, is the only almost exclusively participatory and non-competitive activity.

*Gendered participation differences*: r/Fitness is the only sport topic that yields more than one gender difference term. Female-associated terms covered methods (yoga, pilates), females (woman, girl, female, her), and eating (food, fruit, veggies, eating). The two male terms indicate performance discussions (reps) and same gender interactions (bro).

Table 8. Sports subreddits.

| Rank | Subreddit | Comm. | Fem. | Description | Gendered terms* |
|---|---|---|---|---|---|
| 76 | FantasyFootball | 52434 | 12.6% | Fantasy American football | S: 1 |
| 38 | SquaredCircle | 85930 | 12.6% | Wrestling | S: 1 |
| 69 | Baseball | 57522 | 12.8% | Baseball | 0 |
| 26 | NBA | 131174 | 13.6% | NBA basketball | 0 |
| 75 | MMA | 53409 | 13.7% | Mixed martial arts | 0 |
| 66 | CFB | 64114 | 13.8% | College (American) football | T: 1 ND |
| 24 | NFL | 155396 | 13.9% | American football | S: 1 |
| 50 | Hockey | 73804 | 15.4% | General hockey discussions | 0 |
| 27 | Soccer | 123826 | 15.7% | News, results and discussion | 0 |
| 55 | Sports | 72408 | 16.3% | Sport discussion and news | 0 |
| 35 | Fitness | 89601 | 21.5% | About fitness | T,S: 28 yoga |

*T=Statistically significant evidence of topic differences between genders; S=Statistically significant evidence of style differences between genders. An example gendered topic term is given, when available.

## 5.9 Reposts

The repost categories report content from other sources and are male dominated. Female-associated terms for r/TumblrInAction indicate clothes (wear), gender (trans, girl), and relationships (married).

Table 9. Subreddits reposting from other sources.

| Rank | Subreddit | Comm. | Fem. | Description | Gendered terms* |
|---|---|---|---|---|---|
| 77 | 4Chan | 51947 | 17.0% | Posts from 4Chan | 0 |
| 91 | BestOf | 45857 | 21.8% | Best of Reddit | S: 3 |
| 79 | TumblrInAction | 51202 | 25.0% | Posts from Tumblr | T,S: 7 trans |
| 30 | reddit.com | 106999 | 26.8% | Original reddit, closed in 2012. | T,S: 65 information |

*T=Statistically significant evidence of topic differences between genders; S=Statistically significant evidence of style differences between genders. An example gendered topic term is given, when available.

## 5.10 Personal and domestic advice

Females seem to be more interested in relationship issues (e.g., see: Dunbar, Marriott, & Duncan, 1997), including for news topics (Knobloch-Westerwick & Alter, 2007).

*Gendered participation rates*: There are substantial gender participation differences in the personal and domestic advice subreddits, along largely predictable lines. Two subreddits have mostly female commenters. The r/Relationships and r/Sex subreddits are relatively gender balanced overall, despite the male dominance of Reddit. Discussing relationship issues seems to be a female associated activity, as mentioned above.

*Gendered participation differences*: Most of the subreddits reveal topic gender differences. For r/Relationships, males discuss, or interact with, males more often (dude, bro, man) and females mention females more often (mom, mother), except that they also discuss partners (husband, boyfriend). Other female associated terms include *therapy*, *sorry*, and *together*.

Table 10. Subreddits about personal or domestic advice.

| Rank | Subreddit | Comm. | Fem. | Description | Gendered terms* |
|---|---|---|---|---|---|
| 84 | MaleFashionAdvice | 48124 | 15.3% | Request or give fashion advice | S: 1 |
| 78 | DIY | 51737 | 23.8% | DIY tutorials and help | T,S: 15 mom |
| 74 | AskMen | 54842 | 26.3% | Questions to get a male perspective | T,S: 23 men |
| 31 | LifeProTips | 102922 | 26.5% | Tips that improve life in a specific and significant way | T,S: 48 dog |
| 51 | PersonalFinance | 73109 | 27.3% | Learn how to manage money and invest | T,S: 48 family |
| 45 | Food | 80625 | 29.6% | Food-related posts, preferably with recipes | T,S: 35 recipe |
| 100 | Frugal | 41911 | 34.3% | Frugal  use of resources | T,S: 79 store |
| 58 | Sex | 68987 | 40.2% | Civil discussions about sexuality and sexual relationships | T,S: 149 myself |
| 46 | Relationships | 78391 | 53.2% | Asking specific questions about your relationship | T,S: 138 home |
| 87 | AskWomen | 47217 | 65.0% | Questions to get a female perspective | T,S: 151 night |

*T=Statistically significant evidence of topic differences between genders; S=Statistically significant evidence of style differences between genders. An example gendered topic term is given, when available.

## 5.11 Other

Singling out some topics in this category, an interest in cars seems to be a male trait, perhaps because it is a thing rather than a person (Su, Rounds, & Armstrong, 2009). In contrast, embarrassment seems to be a more frequent social problem for females (Pettijohn, Naples, & McDermott, 2010). Males are more likely to smoke cannabis (e.g., Leatherdale & Ahmed, 2010; Schepis et al., 2011; Tu, Ratner, & Johnson, 2008), and may be more willing to publicly discuss it. Females are more likely to discuss social web images (Thelwall & Vis, 2017).

*Gendered participation rates*: It is unhelpful to compare the gender shares between subreddits in the miscellaneous subreddits discussed here. The male domination of the three image-related subreddits and the r/Cringe subreddit are surprising in the context of the research mentioned above.

*Gendered participation differences*: Six of the ten subreddits reveal topic gender differences. For r/Aww, female-associated terms include baby, cat, kitten, and puppy, whereas male associated terms include joke, ass, gif, and title (of post). For r/TwoXChromosomes, female-associated topics included contraceptives (pill, IUD) and menstruation (period, hormones) and male associated topics included Donald Trump (Trump, republican, government).

Table 11. Unclassified subreddits.

| Rank | Subreddit | Comm. | Fem. | Description | Gendered terms* |
|---|---|---|---|---|---|
| 99 | Cars | 41918 | 15.0% | Vehicle-related discussion, industry news, reviews, projects, videos, DIY guides, art, advice, stories | 0 |
| 86 | Cringe | 47665 | 19.0% | Videos, articles, or personal stories involving an awkward or embarrassing situation | T: 2 her |
| 70 | WoahDude | 57159 | 19.8% | Psychedelic, mind-bending, mesmerizing, reality-distorting, or trippy games, video, audio & images | S: 1 |
| 20 | Trees | 176238 | 21.4% | Cannabis | T,S: 15 pain |
| 36 | GoneWild | 87977 | 22.4% | Pornographic | T,S: 13 bra |
| 42 | OldSchoolCool | 83173 | 24.2% | Pictures and videos of cool kids/teens from history | T,S: 16 grandparent |
| 92 | EarthPorn | 45177 | 24.6% | Landscape photography | S: 7 |
| 72 | Creepy | 56440 | 27.9% | Creepy pictures | T,S: 12 doll |
| 14 | Aww | 291240 | 34.5% | Pictures and videos of cute things | T,S: 171 kitty |
| 53 | TwoXChromosomes | 72771 | 56.2% | Serious and silly content for women's perspectives | T,S: 280 period |

*T=Statistically significant evidence of topic differences between genders; S=Statistically significant evidence of style differences between genders. An example gendered topic term is given, when available.

# 6 Discussion

In answer to the first research question, there are substantial gender differences in the contributors to subreddits. For the most commented 100, although 97 are male dominated, the female proportion varies from 12.6% to 65.0%. The female proportions are similar between subreddits within some of the broad themes (e.g., sport, and technology), giving some confidence that gender compositions are not entirely due to the specifics of the culture within each subreddit.

There are topic-based gender differences within most (58) of the top 100 subreddits and these tend to have many associated terms for subreddits that are more general and with a greater gender balance (RQ2). The number of gender-related terms is large overall (4899, including stylistic terms), but most of these terms were found in the general subreddits. Given the stringent nature of the familywise error rate controlling procedure used, it is nevertheless possible that a statistical hypothesis testing approach that started with hypothesised types of gender difference would be able to detect additional differences compared to the comprehensive approach reported here.

Statistically significant terms for r/Books are summarised here to illustrate the types of insights that can be gained from the results. The r/Books subreddit was chosen for a moderate number of gendered terms and a topic with widely understood (but not well documented) gendered aspects. This topic only produced female-associated terms. Many terms expressed opinions or emotions (e.g., *loved, creepy, beautiful, OMG*), or were stylistic or generic (e.g., *oh, am*), or gender-associated (*husband, boyfriend*), none of which give insights into gendered interests. Several female-associated terms discussed the context in which books were obtained or recommended (e.g., *friend, library, school, daughter*), indicating a greater female interest in the social context of book reading. Gendered interests indicated by the terms include genres (romance, Young Adult (YA)), authors (Jane Austen, Tamora Pierce, Anne McCaffrey, Lois Lowry, Diana Gabaldon, Jacqueline Carey), and books (Anne of Green Gables, The Handmaid's Tail). None of the authors or books would be surprising examples of gendered reading for a book expert, however, except perhaps the ostensibly gender-neutral YA category. The list of statistically significant authors is clearly incomplete, omitting currently extremely popular authors with gendered followings, such as Suzanne Collins (The Hunger Games series) and Stephenie Meyer (the Twilight series). Thus, the gendered interests discovered are a *sample* rather than a comprehensive set of the most gendered issues for books.

From the perspective of gendered interests, r/TIFU illustrates that gendered topics may not be "interests" in the sense of topics that someone might describe as being an interest. This subreddit is recreational but the gendered terms point to aspects of situations that caught their attention but not something that the user would regard as "an interest". For example, whilst the topics *cats* (cat, kitty), injuries (hurt, doctor) and hair might be female-associated interests, the topic *tampon* is about a necessity for many women rather than a recreational interest.

Reddit could be used to infer underlying gender differences about the topics discussed (RQ3). For this, an important limitation is that many factors influence the gender composition of a subreddit, including the following.

- Gendered interest in the underlying topic.
- Gendered interest in 'writing oneself into being' to participate (e.g., see Chapter 4 of: Boyd, 2008).
- Gendered interest in the way that the topic is approached in the subreddit.
- The overall gender composition of Reddit, on the basis that a person is more likely to find and use a subreddit if they are already a member of Reddit.
- Whether the subreddit has been an automatic subscription or a hot topic.
- External factors, such as associated podcasts, websites, or organisations.
- Competition within Reddit from other subreddits with overlapping themes and a different gender perspective (e.g., r/AskWomen vs. r/AskReddit).
- Moderator policies and commenting cultures.
- Age differences in gendered interests for the topic.
- International differences in gendered interests for the topic – especially for countries that use Reddit most.
- Changes in gendered interests in a topic during the lifetime of Reddit.

Because of the above factors, the share of females for a subreddit gives a biased and unreliable estimate of the underlying female interest in the topic discussed. Thus, whilst the gender share of comments posted to a subreddit gives an indication of the overall level of female interest in the topic, it is likely to be male biased (because Reddit is) and to have competing biases to an extent that it is not possible to calculate a corrected estimate of the underlying level of gendered interest in the topic. The substantial impact that these factors can have is clear from the examples above of male-dominated subreddits with gender neutral or female-oriented topics, such as r/Aww and r/books). Since influences can vary between subreddits, it is also not possible to conclude that one subreddit's topic is more female-friendly because it has a larger proportion of female commenters.

Despite the above limitations, social science research is rarely conducted in experimental conditions where influencing factors can be controlled. The limitations on Reddit, whilst substantial, may not be worse than, for example, low response rate self-selection survey biases, or the biases that are inevitable within small sample qualitative research using interviews or group discussions and these are nevertheless accepted and valuable research approaches (Martin, 1983; Merriam, & Tisdell, 2015).

An unexpected problem with the gender word association tests was that females tended to write comments that were longer, on average (the opposite has been observed in the similar context of newsgroups: Sussman & Tyson, 2000). The main cause of this was a higher proportion of very short (up to ten words) posts from males. The length disparity gave an advantage to females in the word association tests since a higher proportion of female-authored comments were substantial contributions. This resulted in there being more significant terms for females than for males in the word association tests for each subreddit, and often no terms for males.

An issue that reduces the statistical reliability of the word association results, despite the steps taken to control this, is that sharing and imitation is common in social web sites and so the word frequency data violates the independence assumptions needed to interpret the data. For example, the statistical significance of the term “doll” in r/Creepy may be due to one widely imitated post or many comments on a single post with a doll theme that incorporated an aspect of interest to females.

Finally, for a quantitative analysis of patterns, eliminating too many comments from a single poster was necessary, although this reduces the effective sample size of Reddit to a few million comments. Whilst this is still a large sample, it may only be enough for investigating topics that are widely discussed within Reddit. Thus, in practice, only a small percentage of Reddit comments can be investigated quantitatively for gender differences.

# 7 Conclusions

Reddit is a free source of billions of social web comments that can be investigated for insights into the topics discussed. This article focused on gender, which was detectable from 5% of comments. The widespread use of Reddit makes its perspective on gender important for researchers needing a more complete picture of online communication as well as for insights into gendered interests in topics. In this investigation, 95.1% of the comments had to be discarded because the author gender could not be inferred from their name. After removing multiple comments from the same user in a single month, only 1% of the original Reddit comments were usable for gender analyses. Nevertheless, Reddit is relatively straightforward to access and analyse with free software (code created for this article is shared through mozdeh.wlv.ac.uk). Whilst the analysis here focused on topics defined by subreddits, Reddit could instead be filtered through keyword matching or other methods to get other sets of comments to analyse.

The results showed differences in the proportions of male and female subreddit commenter genders between topics (and broad themes), but these cannot be safely generalised to evidence of gendered interests in subreddit topics due to the range of factors that affect participation rates, as discussed above. Thus, whilst gender participation rates vary greatly between subreddits (RQ1), this cannot be used as strong evidence of gender differences in interests (RQ3), but only as indicative of gender differences, taking into account the factors that affect participation rates.

Statistically significant gender differences in topic-related terms within comments were detected for most of the 100 most commented subreddits (RQ2). Thus, the results show that Reddit can be used to suggest gender differences in topic choice and for evidence of statistically significant gender differences in contributions to topics. The statistical evidence is not robust enough to infer underlying gender differences in interests (RQ3) because imitation and sharing in social web sites violates the independence assumption of the tests used, an issue that is impossible to circumvent. The likelihood of a small number of causes driving a significant result could be checked qualitatively by reading a sample of posts containing a keyword. If this strategy is accepted then RQ3 would have a positive answer. This direct checking is not possible for topic modelling because of the multidimensional relationship between topics and documents. Full sets of terms extracted and significance levels are available in the online appendix for researchers and practitioners interested in gender differences to consult (doi:10.6084/m9.figshare.6470930).

The current paper employed the cautious Benjamini-Hochberg approach to reduce the risk of false positive results. If Reddit is used to get insights into a topic that would be followed up by other methods (e.g., interviews, surveys) then this could be relaxed, giving a much greater number of terms to investigate, some of which are likely to be relevant. Similarly, hypothesis-driven research could also be statistically more powerful by focusing on a set of terms to analyse rather than (as above) analysing all terms with a high significance threshold to protect the familywise error rate. Thus, the practical conclusion of this article is that Reddit is suitable for research into gendered issues discussed in the site with a hypothesis-driven approach or as part of a multi-methods research approach. This includes theoretical social science investigations or commercial studies to understand the role of gender for a given product segment.